\documentclass[aps,prl,twocolumn,showpacs,superscriptaddress]{revtex4-1}
\usepackage{graphicx}
\usepackage{amsmath}
\usepackage{amssymb}
\usepackage{colordvi}
\usepackage{mathrsfs}
\usepackage{bm}
\usepackage{verbatim}
\usepackage{dcolumn}
\usepackage{bm}
\usepackage{epsfig}
\usepackage{subfigure}
\usepackage{dsfont}

\usepackage[colorlinks,linkcolor=blue,anchorcolor=blue,citecolor=blue]{hyperref}

\begin{document}
\title{Second-Order Topological Insulator in van der Waals Heterostructures of CoBr$_2$/Pt$_2$HgSe$_3$/CoBr$_2$}
\author{Zheng Liu}
\affiliation{CAS Key Laboratory of Strongly-Coupled Quantum Matter Physics, and Department of Physics, University of Science and Technology of China, Hefei, Anhui 230026, China}
\author{Yafei Ren}
\affiliation{Department of Materials Science and Engineering, University of Washington, Seattle,  Washington 98195, USA}
\author{Yulei Han}
\affiliation{Department of Physics, Fuzhou University, Fuzhou, Fujian 350108, China}
\affiliation{CAS Key Laboratory of Strongly-Coupled Quantum Matter Physics, and Department of Physics, University of Science and Technology of China, Hefei, Anhui 230026, China}
\author{Qian Niu}
\affiliation{CAS Key Laboratory of Strongly-Coupled Quantum Matter Physics, and Department of Physics, University of Science and Technology of China, Hefei, Anhui 230026, China}
\author{Zhenhua Qiao}
\email[Correspondence author:~~]{qiao@ustc.edu.cn}
\affiliation{CAS Key Laboratory of Strongly-Coupled Quantum Matter Physics, and Department of Physics, University of Science and Technology of China, Hefei, Anhui 230026, China}
\affiliation{ICQD, Hefei National Laboratory for Physical Sciences at Microscale, University of Science and Technology of China, Hefei, Anhui 230026, China}

\date{\today{}}

\begin{abstract}
  Second-order topological insulator, which has (d-2)-dimensional topological hinge or corner states, has been observed in three-dimensional materials, but has yet not been observed in two-dimensional system. In this Letter, we theoretically propose the realization of second-order topological insulator in the van der Waals heterostructure of CoBr$_2$/Pt$_2$HgSe$_3$/CoBr$_2$. Pt$_2$HgSe$_3$ is a large gap $\mathbb{Z}_2$ topological insulator. With in-plane exchange field from neighboring CoBr$_2$, a large band gap above 70 meV opens up at the edge. The corner states, which are robust against edge disorders and irregular shapes, are confirmed in the nanoflake. We further show that the second-order topological states can also be realized in the heterostructure of jacutingaite family $\mathbb{Z}_2$ topological insulators. We believe that our work will be beneficial for the experimental realization of second-order topological insulators in van der Waals layered materials.
\end{abstract}

\maketitle

\textit{Introduction---.} The second-order topological insulator~\cite{2013_Fan,2017_Benalcazar,2017_PRB_Benalcazar,2018_Ezawa,2020_R_Chen,2020_Chun_Bo_Hua,2020_Agarwala,2018_Schindler,2017_Langbehn,2017_Song,2018_Geier,2018_Khalaf,2018_Miert,2019_Benalcazar,2018_Kunst,2020_YB_Yang,2020_QB_Zeng,2018_Ezawa2,2019_F_Liu,2019_Calugaru,2019_K_Kudo,2020_Ren,2020_Rui_Xing,2019_Trifunovic,2020_Rasmussen} is a kind of topological state of matter that possess 0-dimensional (0D) corner or 1D hinge states for 2D or 3D system, respectively. Since it was first conceptually proposed~\cite{2017_Benalcazar,2017_PRB_Benalcazar}, second-order topological insulators have been widely studied in the aspects of lattices~\cite{2018_Ezawa,2020_R_Chen,2020_Chun_Bo_Hua,2020_Agarwala}, symmetries~\cite{2018_Schindler,2017_Langbehn,2017_Song,2018_Geier,2018_Khalaf,2018_Miert,2019_Benalcazar}, model constructions~\cite{2018_Kunst,2020_YB_Yang,2020_QB_Zeng,2018_Ezawa2,2019_F_Liu,2019_Calugaru,2019_K_Kudo,2020_Ren,2020_Rui_Xing}, and topological classifications~\cite{2019_Trifunovic,2020_Rasmussen}. Inspired by these proposals, some potential applications of second-order topological insulators were proposed~\cite{2020_Banerjee,2021_chang_An}. So far, second-order topological insulators have only been experimentally realized in 3D materials, i.e., bismuth~\cite{2018_Schindler2}, Bi$_4$Br$_4$~\cite{2021_Noguchi}. In 2D, material candidates of second-order topological insulators are still limited~\cite{2019_Sheng,2020_E. Lee,2019_Bing_Liu,2021_Cong_Chen,2019_Park,2021_Bing_Liu,2020_Cong Chen}, and the material realization of second-order topological insulators in electronic systems is still rare, which greatly limits the potential development of this field. Therefore, it is highly desirable to explore new material candidates and scalable methods for the 2D second-order topological insulator.

To design second-order topological state, breaking specified symmetry in first-order topological insulators is a scalable scheme~\cite{2020_Ren}, which is believed to be easily implemented by applying external pressure or introducing magnetization. Recently, some predictions have been made in 3D systems such as SnTe~\cite{2018_Schindler}, EuIn$_2$As$_2$~\cite{2019_Yuanfeng Xu} and Sm-doped Bi$_{2}$Se$_3$~\cite{2019_Yue}. In 2D system, the only prediction is made in bismuthene deposited on a bulk magnetic insulator that provides in-plane magnetization as the symmetry breaking term~\cite{2020_Cong Chen}. However, no experimental progress has been made in this material system. Compared to the heterostructure of bulk magnetic substrate, van der Waals (vdW) heterostructure~\cite{2013_Geim,2016_Novoselov} constructed by 2D magnetic layers can avoid the cleaved surface problem, which makes it experimentally friendly.

\begin{figure}
  \includegraphics[width=8cm,angle=0]{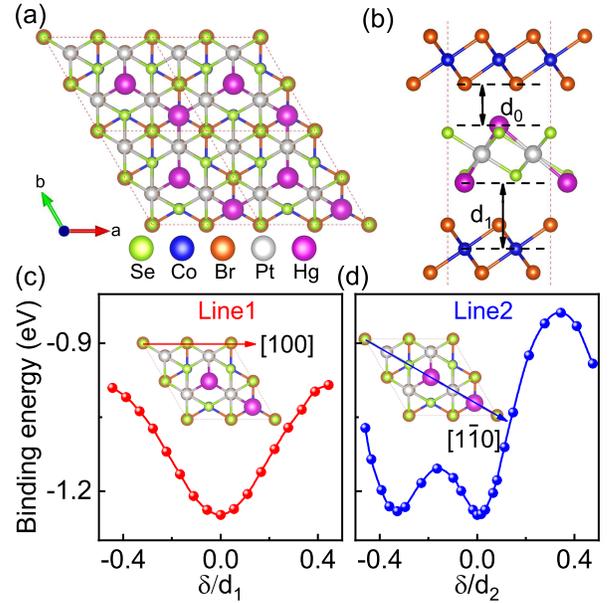}
  \caption{(a)-(b) Top and side views of the heterostructures of CoBr$_2$/Pt$_2$HgSe$_3$/CoBr$_2$ . (c)-(d) Binding energies along the high symmetry lines for the respective horizontal and diagonal directions. Here, $\delta$ represents the displacement, d$_1$ and d$_2$ are half the length of the lattice constants of horizontal and diagonal directions.}
  \label{Fig1}
\end{figure}

In this Letter, we show the possibility of realizing the second-order topological insulators in the vdW heterostructure of CoBr$_2$/Pt$_2$HgSe$_3$/CoBr$_2$ with large band gap and robust corner states. In the absence of spin-orbit coupling (SOC), the magnetic proximity effect leads to a large spin splitting over 200 meV at valleys K and K$'$. When the SOC is considered, a sizeable bulk band gap of 79.7 meV and nanoribbon band gap of 72.8 meV are opened. In the nanoflake, one topological corner state arises at the intersection of boundaries with its eigenenergy being locating inside the energy gap of edge states. When irregular boundary and Anderson disorders are introduced, we show that the topological corner states are almost unaffected. Besides Pt$_2$HgSe$_3$, we find that other $\mathbb{Z}_2$ topological insulators of the jacutingaite family can also be utilized as the candidate materials for the realization of topological corner states.
Moreover, a low-energy effective model based on topological edge states is constructed, demonstrating that the 1D Jackiw-Rebbi model can be used to explain the presence of topological corner states.

\textit{Atomic Structure and Calculation Methods---.} Figures~\ref{Fig1}(a) and ~\ref{Fig1}(b) display the heterostructure of CoBr$_2$/Pt$_2$HgSe$_3$/CoBr$_2$, where 3D bulk Pt$_2$HgSe$_3$ is a dual-topological semimetal that can be exfoliated down to a few layers in ambient conditions~\cite{2019_Ghosh,Exp_PHS1,Exp_PHS2}, monolayer Pt$_2$HgSe$_3$ is a $\mathbb{Z}_2$ topological insulator with large band gap of 0.17 eV~\cite{DFT_PHS1,DFT_PHS2}, and monolayer CoBr$_2$ is a ferromagnetic insulator with an in-plane magnetic easy axis~\cite{DFT_CoBr2}. In our study, we adopt the 1$\times$1 Pt$_2$HgSe$_3$ and 2$\times$2 CoBr$_2$ supercells with a lattice mismatch of $\sim$3.4\%. The structural stability is strictly checked by calculating the binding energy of Pt$_2$HgSe$_3$/CoBr$_2$ heterostructure with a series of different stacking configurations. The binding energy can be expressed as $\Delta E=E_{\text{H}}-E_{\text{P}}-E_{\text{C}}$, where $E_{\text{H}}$, $E_{\text{P}}$ and $E_{\text{C}}$ are respectively the total energy of the heterostructure, Pt$_2$HgSe$_3$ monolayer, and CoBr$_2$ monolayer. Figures~\ref{Fig1}(c) and \ref{Fig1}(d) display the calculated binding energies for Pt$_2$HgSe$_3$/CoBr$_2$ moving along [100] and [1$\overline{1}$0] directions, respectively. The system illustrated in Figs.~\ref{Fig1}(a) and \ref{Fig1}(b), owning the lowest binding energies, is the most stable structure.

\begin{figure}
  \includegraphics[width=8.5cm,angle=0]{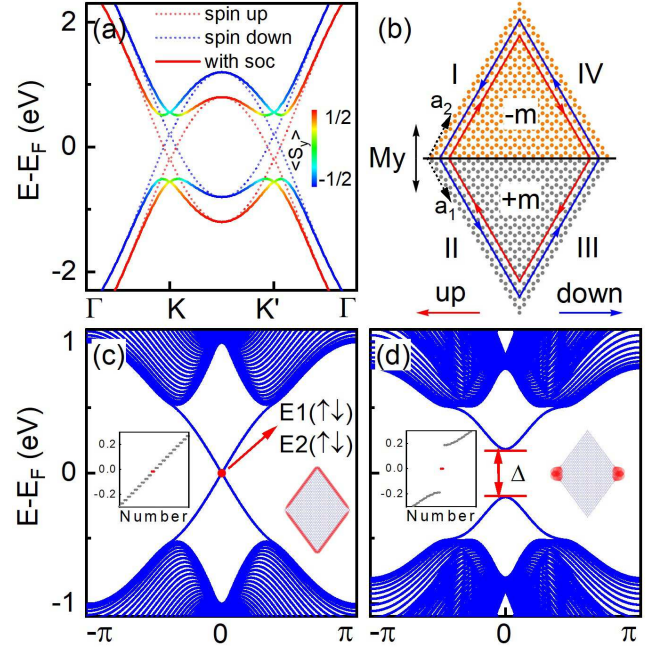}
  \caption{(a) The band structure of in-plane magnetized hexagonal lattice without(dashed line) and with(solid line) SOC. The color of solid line represents the expectation value of s$_y$ operator. (b) The schematic diagram of spin up/down edge states in Kane-Mele type topological insulator and mass term when in-plane magnetization is induced. (c) and (d) show the band structure of a zigzag nanoribbon without(c) and with(d) in-plane magnetization, respectively. The insets in (c) and (d) shows the energy spectrums of 0 D nanoflake and probability distribution of the states marked in red.}
	\label{Fig2}
\end{figure}

Our first-principles calculations were performed by using the projected augmented-wave method~\cite{PAV} as implemented in the Vienna \textit{ab initio} simulation package (VASP)~\cite{VASP}. The generalized gradient approximation of the Perdew-Burke-Ernzerhof type was used to describe the exchange-correlation interaction~\cite{PBE}. All atoms were allowed to relax until the Hellmann-Feynman force on each atom is smaller than 0.01 eV/\AA. The $\mathrm{\Gamma}$-centered Monkhorst-Pack grid of $\mathrm{7\times 7\times 1}$ was carried out in all our calculations. For Co element, the GGA+$U$ method was used with the on-site repulsion energy $U=3.67~\mathrm{eV}$~\cite{DFT_CoBr2}. The vdW interaction was treated by using DFT-D2 functional~\cite{DFT-D2}. And the topological properties were calculated by using maximally-localized Wannier functions as implemented in Wannier90 package~\cite{wannier90}. A vacuum buffer layer of 20 {\AA} was used to avoid interaction between adjacent slabs. The plane-wave energy cutoff was set to be 400 eV.

\textit{Model Analysis---.}
Before demonstrating detailed first-principles calculation results, it is necessary to clearly illustrate the underlying physics. The physical model of the heterostructure can be described by the Kane-Mele model with in-plane magnetization~\cite{DFT_PHS1,DFT_PHS2,DFT_PHS3,2021_zheng,Kane-Mele model}, which can be expressed as
\begin{eqnarray}\label{eq:Ham}
H & = & t\sum_{\langle ij \rangle \alpha}c_{i\alpha}^{\dagger}c_{j\alpha}+i\lambda_{\rm ISO}\sum_{\langle\langle ij \rangle\rangle\alpha\beta}v_{ij}s^{\alpha\beta}_{z}c_{i\alpha}^{\dagger}c_{j\beta} \notag\\
&& +\sum_{i\alpha\beta}m_0s^{\alpha\beta}_yc_{i\alpha}^{\dagger}c_{i\beta},
\end{eqnarray}
where $c^{\dagger}_{i\alpha}$($c_{i\alpha}$) is the creation (annihilation) operator for electron on site $i$ with spin $\alpha$. The first term is the nearest neighbor hopping term with hopping amplitude $t$. The second term represents the intrinsic SOC that involves the next-nearest neighbor hopping with an amplitude of $\lambda_{\text{ISO}}$. $v_{ij}=\boldsymbol{d}_j\times\boldsymbol{d}_i/|\boldsymbol{d}_j\times\boldsymbol{d}_i|$, where $\boldsymbol{d}_i$ and $\boldsymbol{d}_j$ are two nearest neighbor bonds connecting the next-nearest neighbor sites. The third term is the in-plane magnetization along $y$ direction.
In the presence of in-plane magnetization, the bulk band structure becomes split upward (downward) with the $s_y$ eigenvalue equals +1 (-1), respectively, as displayed in dashed lines of Fig.~\ref{Fig2}(a). When the intrinsic SOC is further included, four anticrossings occur at the bands with opposite spin directions as the intrinsic SOC can mix up the $s_y=\pm1$ eigenstates.

To clearly understand the second-order topological state, we construct a low-energy effective model on the basis of the edge states of $\mathbb{Z}_2$ topological insulator that gives the edge-corner correspondence, just like the bulk-edge correspondence in first-order topological insulators. For simplicity, we first construct the effective model for the edge states of zigzag nanoribbon that is periodic along $a_1$ direction and has two edges along $a_2$ direction. As illustrated in Fig.~\ref{Fig2}(b), the two edges of the nanoribbon are marked as II and IV.
On edge II, the basis functions of the edge states are $|E_1,\uparrow\rangle$ and $|E_1,\downarrow\rangle$ that can be obtained numerically.
In the absence of in-plane magnetization, both states are decoupled as they form a Kramers pair and the low-energy effective model of the linear dispersion can be expressed as $H_{\text{edge}}=-\eta v_Fk_1\sigma_z$, with $\eta=+1$. In the presence of an in-plane magnetization that breaks the time reversal symmetry, the edge modes are coupled and the Hamiltonian becomes $H_{\text{II}}=-\eta (v_Fk_1\sigma_z-m\sigma_x)$. The coupling matrix elements are obtained by using the numerical basis functions.
Along edge IV, the velocity of spin-up (down) electron is opposite to that at edge II, as highlighted by the arrows in Fig.~\ref{Fig2}(b), which manifests the nature of the $\mathbb{Z}_2$ topological insulator.
Therefore, the edge Hamiltonian reads $H_{\text{IV}}=-\eta (v_Fk_1\sigma_z-m\sigma_x)$ with $\eta=-1$ indicating the velocity reverse. The term $m\sigma_x$ from magnetization opens a band gap $\Delta=2m$ at the edges. A similar analysis can be carried out for the zigzag nanoribbon that is periodic in $a_2$ direction, and the effective model becomes $H_{\text{I,III}}=\eta (v_Fk_2\sigma_z-m\sigma_x)$, with $\eta=+1 (-1)$ for edge I (III). To be more transparent, the effective model can be rewritten by taking ``edge coordinate" $l$ that grows anticlockwisely,
\begin{eqnarray}\label{eq:EffHam}
H_{\text{eff}} =iv_F\sigma_z\partial_l+m(l)\sigma_x,
\end{eqnarray}
where $m(l)=+m (-m)$ for edge II and III (I and IV). By applying an unitary transformation $U=\exp(i\sigma_y{\pi}/{4})$, Eq.~(\ref{eq:EffHam}) becomes
\begin{eqnarray}\label{eq:EffHam_2}
H^\prime_{\text{eff}} =-iv_F\sigma_x\partial_l+m(l)\sigma_z,
\end{eqnarray}
which is exactly the 1D Jackiw-Rebbi model~\cite{Jackiw-Rebbi}. Thus, there always exist zero energy solutions near the domain walls, where $m$ changes its sign. The numerical results are consistent with our effective model. As displayed in Figs.~\ref{Fig2}(c) and \ref{Fig2}(d), the edge spectrum is gapped in the presence of in-plane magnetization. When considering a nanoflake, we can observe two zero-energy states with wavefunction distributed at the corners.

\begin{figure}
  \includegraphics[width=8.5cm,angle=0]{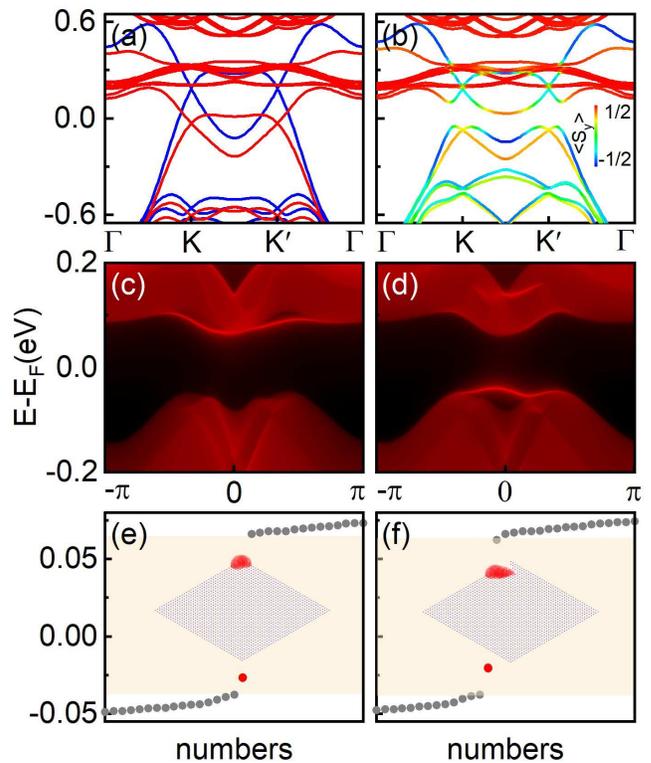}
  \caption{(a)-(b) Band structure of CoBr$_2$/Pt$_2$HgSe$_3$/CoBr$_2$ heterostructure without and with SOC. In (a), the red (bule) represents spin up (down) states. The color of line represents the the expectation value of s$_y$ operator in (b). (c)-(d) Edge states of zigzag nanoribbon with (c) and (d) representing the right and left terminals, respectively. (e)-(f) Energy levels of the nanoflake. Corner states are high-lighted in red. The insets show the distribution of the corner state. In (f) the corner state still exist with irregular boundaries.}
  \label{Fig3}
\end{figure}

\textit{Band Structures and Second-Order Topological Properties---.}
The band structure evolution from first-principles calculations agrees well with our model analysis. As reported in previous studies~\cite{DFT_PHS1,DFT_PHS2}, Pt$_2$HgSe$_3$ is a Kane-Mele type topological insulator, which exhibits two Dirac cones at K and K$'$ as graphene in the absence of SOC and opens a large band gap at the Dirac points after considering SOC. When CoBr$_2$ cover layers are introduced, the band structure of Pt$_2$HgSe$_3$ is obviously modified. In the absence of SOC, the spin majority and spin minority bands of Pt$_2$HgSe$_3$ are largely separated by 211.1 meV at K/K$'$ point (see Fig.\ref{Fig3}(a)), indicating a strong magnetic proximity effect in this sandwiched structure. When the SOC is further considered, large band gaps open up around the band crossing points as illustrated in Fig.~\ref{Fig3}(b). The spin projections $\langle s_y \rangle$ on the band structures are also provided, and the spin is mixed around the Fermi level, demonstrating that the gaps are opened by spin mixing effect.

\begin{figure}
  \includegraphics[width=8.5cm,angle=0]{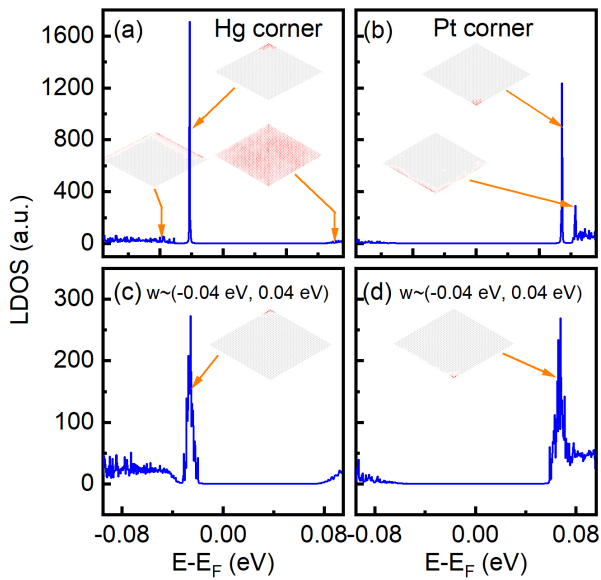}
  \caption{(a)-(b) Local density of states at Hg and Pt corners, respectively. Insets show the local density of states at real space for corner, edge and bulk states at specific energy. (c)-(d) The average local density of states in the presence of disorder for Hg (c) and Pt (d) corners, respectively. Disorder strength is set as $W=80$ meV. Over 100 samples are collected.}
  \label{Fig4}
\end{figure}

To explore the topological properties of the heterostructure, the energy spectra of 1D nanoribbon and 0D nanoflake are calculated by using the Hamiltonian generated from the maximally localized Wannier functions~\cite{wannier90}. The atomic orbitals of Hg (s) and Pt ($d_{xy}$, $d_{yz}$, $d_{x^2}$, $d_{yz}$) are used for projection since they contribute dominantly to the energy bands near the Fermi level. In Figs.~\ref{Fig3}(c) and \ref{Fig3}(d), we plot the edge states of 1D zigzag nanoribbon by using the surface Green's function technique. Large band gaps of 72.8 meV are opened at the edges. To verify the formation of corner states, we calculate the energy spectrum of the nanoflake system with a $40\times40$ unit cell. As shown in Fig.~\ref{Fig3}(e), we find one in-gap state highlighted in red, with its probability density distributed around one corner. When disorders are introduced by introducing edge randomness, we find that the topological corner state still exists at the irregular region [see Fig.~\ref{Fig3}(f)].

One may observe that the topological corner states in Pt$_2$HgSe$_3$ heterostructure are slightly different from those in the model Hamiltonian. The reason is that the atoms on the upper edges are Hg while the atoms on the lower edges are Pt, which leads to different onsite potential on the opposite edges.
As a result, one of the corner states still lies inside the band gap while the other is outside of the energy gap. This corner state can be tuned into the gap when applying a negative onsite potential to the edges formed by Pt atoms.
In Supplemental Material~\cite{SM}, we demonstrate that the second-order topology will not change as long as the topological corner states exist. For 1D nanoribbon, the edge states are moved up/down when onsite potentials are considered. For 0D nanoflake, when different onsite potentials are introduced, the degeneracy of two corner states is broken and the wavefunctions of two corner states are distributed at different corners.

\textit{Robustness of the Corner States---.} To explore the robustness of the corner states, we add random disorders $H_d$ to the outermost unit cells, where $H_d=w\sum_ic_{i}^{\dagger}c_{i}$ with $w$ being uniformly distributed within an interval of $[-W/2,W/2]$. The disorder strength is set as $W=80$ meV that is approximately in the same magnitude of edge band gap. The local density of states (LDOS) around the corner is introduced to characterize the existence of corner states. The LDOS can be calculated by using retarded Green's function
\begin{eqnarray}\label{eq:EffHam_3}
\text{LDOS}(E,n)=-\frac{1}{\pi}\text{Im}\big[\frac{1}{E-H+i\delta}\big]_{nn},
\end{eqnarray}
where $n$ represents the atomic site. Figure~\ref{Fig4} plots the LDOS summation of atomic sites at three unit cells around the corners with obtuse angles. In the absence of disorder, sharp peaks occur in Fig.~\ref{Fig4}(a) and (b), corresponding to corner states at Hg and Pt edges respectively. Aside from the typical peaks, LDOS can also provide real-space distributions of electronic states. The electronic states are predominantly localized around the corner at the peaks of the LDOS curve, as illustrated in the inset of Fig.~\ref{Fig4}(a) and (b). When the random disorders are introduced, we plot the averaged LDOS on 100 samples as displayed in Figs.~\ref{Fig4}(c) and (d). One can see that the peaks for corner states are still visible, suggesting that topological corner states are robust against weak disorder.

\begin{table}
	\caption{Structural, band and topological properties of heterostructures in Pt$_2$HgSe$_3$ family.}
	\begin{ruledtabular}
		\begin{tabular}{ccccc}
			Heterostructure&Lattice Mismatch&Band gap&2${nd}$-order TI \\ \hline
			Pt$_2$ZnS$_3$/CoBr$_2$&4.01 \%  &8.3 meV&No \\
			Pt$_2$ZnS$_3$/CoCl$_2$&1.37 \%  &29.4 meV&No \\
			Pt$_2$ZnS$_3$/NiBr$_2$&1.68 \%  &0 meV&No \\
			Pt$_2$ZnS$_3$/NiCl$_2$&3.14 \%  &24.1 meV&Yes \\
			Pt$_2$HgS$_3$/CoBr$_2$&3.78 \%  &58.0 meV&Yes \\
            Pt$_2$HgS$_3$/CoCl$_2$&1.60 \%  &88.6 meV&Yes \\
            Pt$_2$HgS$_3$/NiBr$_2$&1.45 \%  &62.9 meV&No \\
            Pt$_2$HgS$_3$/NiCl$_2$&3.37 \%  &48.9 meV&No \\
		\end{tabular}
	\end{ruledtabular}
	\label{table-1}
\end{table}

\textit{Corner states in Pt$_2$HgSe$_3$ Family---.} To explore the possibility of realizing corner states in other heterostructures of Pt$_2$HgSe$_3$ family materials~\cite{DFT_PHS2,2020_Lima,2020_Ma}, we systematically study the electronic band structures and topological properties of MZ$_2$/Pt$_2$XS$_3$/MZ$_2$ (M = Co and Ni; Z = Br and Cl; X = Zn and Hg). As displayed in Table~\ref{table-1}, most heterostructures have small lattice mismatch and sizable band gaps. By calculating the energy spectrums and wavefunction distributions of nanoflake, three candidates with topological corner states are discovered (see details in Supplemental Materials~\cite{SM}).

\textit{Summary---.} We demonstrate that the two-dimensional second-order topological states can be realized in the vdW heterostructures of CoBr$_2$/Pt$_2$HgSe$_3$/CoBr$_2$. CoBr$_2$ layers proximity-induce a considerable in-plane exchange field in Pt$_2$HgSe$_3$, which makes the edge states gapped. Inside the band gap, we find corner states in a nanoflake geometry that can be understood by a 1D Jackiw-Rebbi model. We find that the corner states can be probed by measuring the local density of states near the corner, which is robust against the atomic randomness at the boundaries and Anderson disorders. We show that this topological corner states can also be realized in other candidate materials, e.g., MZ$_2$/Pt$_2$XS$_3$/MZ$_2$ (M = Co and Ni; Z = Br and Cl; X = Zn and Hg), which should be beneficial to the experiment observations of corner states in electronic systems.

\begin{acknowledgments}
This work was financially supported by the National Natural Science Foundation of China (11974327 and 12004369), Fundamental Research Funds for the Central Universities (Grants No. WK3510000010 and WK2030020032), and Anhui Initiative in Quantum Information Technologies (Grant No. AHY170000). We are grateful to AMHPC and Supercomputing Center of USTC for providing the high-performance computing resources.
\end{acknowledgments}

\end{document}